\documentclass[11pt,onecolumn]{IEEEtran}
\usepackage{graphicx,subfigure}
\usepackage{amsfonts,amsmath,amssymb, mathrsfs}
\usepackage{multirow}
\usepackage{dsfont}
\usepackage{epic,eepic,eepicemu}
\usepackage{epsf}
\usepackage{epsfig}
\usepackage{graphics}
\usepackage{psfrag}
\usepackage{tikz}
\usepackage{url}
\usepackage{enumerate}

\newtheorem{theorem}{Theorem}
\newtheorem{definition}{Definition}

\newtheorem{remark}{Remark}
\def\cxe{\mathcal{K}}
\DeclareMathOperator{\Ber}{Ber}

\usepackage{xspace}
\usepackage{mathrsfs}
\usepackage{amsopn}

\newcommand{\Sc}{\mathcal{S}}

\def\la{\lambda}

\newcommand{\E}{\mathbb{E}}

\def\textiid{i.i.d.\@\xspace}
\newcommand\iid{\ifmmode\text{ i.i.d. } \else \textiid \fi}

\newcommand{\beqs}{\begin{equation*}}
\newcommand{\eeqs}{\end{equation*}}
\newcommand{\beq}{\begin{equation}}
\newcommand{\eeq}{\end{equation}}

\begin{document}
\title{On Maximal Correlation, Hypercontractivity, and the Data Processing 
Inequality studied by Erkip and Cover}

\author{Venkat Anantharam, Amin Gohari,  Sudeep Kamath, and Chandra
  Nair}
\author{
    \IEEEauthorblockN{Venkat Anantharam\IEEEauthorrefmark{1}, Amin Gohari\IEEEauthorrefmark{2}, Sudeep Kamath\IEEEauthorrefmark{1}, Chandra Nair\IEEEauthorrefmark{3}}\\ 
    \IEEEauthorblockA{\IEEEauthorrefmark{1}EECS Department, University of California, Berkeley,
    \\\{ananth, sudeep\}@eecs.berkeley.edu}\\
    \IEEEauthorblockA{\IEEEauthorrefmark{2}EE Department, Sharif University of Technology, Tehran, Iran
    \\aminzadeh@sharif.edu}\\
\IEEEauthorblockA{\IEEEauthorrefmark{3} IE Department, The Chinese University of Hong Kong\\
chandra@ie.cuhk.edu.hk}
}

\maketitle

\begin{abstract}
  In this paper we provide a new geometric characterization of the
  Hirschfeld-Gebelein-R\'{e}nyi maximal correlation of a pair of random
  $(X,Y)$, as well as of the chordal slope of the nontrivial boundary
of the hypercontractivity ribbon of $(X,Y)$ 
at infinity. The new characterizations lead to simple proofs for some of the known 
facts about these quantities. We also
  provide a counterexample to a data processing inequality
claimed by Erkip and Cover, and find the correct tight constant for this
kind of inequality.
\end{abstract}

\section{Introduction}
There are various measures available to quantify the dependence between two random
variables. A well-known such measure for real-valued random variables
is the Pearson correlation coefficient
$\rho_p(X,Y) :={\mathrm{cov}(X,Y) \over \sigma_X \sigma_Y}$, which quantifies
the linear dependence between the two random variables. A closely related
measure, called the Hirschfeld-Gebelein-R´enyi maximal correlation, or
simply the maximal correlation, measures the cosine of the angle between the linear
subspaces of mean zero square integrable real-valued random variables defined
by the individual random variables, as below. 

\begin{definition}
Given random variables $X$ and $Y$, 
the Hirschfeld-Gebelein-R\'{e}nyi maximal correlation of $(X,Y)$ is defined as follows:
\begin{equation}\rho_m(X;Y) :=\max_{(f(X),g(Y)) \in \Sc} \E[f(X)g(Y)],\label{eq:1}\end{equation}
where $\Sc$ is the collection of pairs of real-valued random variables $f(X)$ and $g(Y)$
such that
$$ \E f(X) = \E g(Y) = 0, ~\mbox{and}~\E f^2(X) = \E g^2(Y) = 1.$$
If $\Sc$ is empty
(which happens precisely when at least one of $X$ and $Y$ is constant almost surely) 
then one defines $\rho_m(X;Y)$ to be $0$.
\hfill $\Box$
\end{definition} 

This measure,
first introduced by Hirschfeld \cite{Hirschfeld} and Gebelein
\cite{Gebelein} and then studied by R\'{e}nyi \cite{Renyi}, has found
interesting applications in information theory. 

As a general remark, to stay clear
of technicalities,
we restrict ourselves throughout this paper 
to discrete random variables $(X,Y)$ taking values in $\mathcal{X} \times \mathcal{Y}$ with $|\mathcal{X}|, \mathcal{Y}| < \infty$. Further we assume that
$\mathbb{P}(X=x) > 0, \forall x \in \mathcal{X}$ and $\mathbb{P}(Y=y)>0~\forall y, \in \mathcal{Y}$. We will use 
$:=$ and occasionally $=:$ for equality by definition.

\begin{definition}
For any real-valued random variable $X$ and real number
$\mathsf{p}\neq 0$, define $\|X\|_\mathsf{p} := (\E |X|^\mathsf{p})^{\frac {1}{\mathsf{p}}}$. Define
$\|X\|_0 := \exp(\E(\log|X|)).$ For $\mathsf{p} \leq 0$, $\|X\|_\mathsf{p}= 0$ if
$\mathbb{P}(|X| = 0) > 0$.
\hfill $\Box$
\end{definition}

Renyi \cite{Renyi} derived an alternate characterization to $\rho_m(X,Y)$ as follows:
\begin{equation}
\rho_m(X;Y)=\max_{f(X): \E f(X)=0,
    \E[f^2(X)]=1}\|\E[f(X)|Y]\|_2.\label{eq:2i}\end{equation}

Maximal correlation has interesting connections to the
hypercontractivity of Markov operators, as demonstrated by Ahlswede
and G\'{a}cs in \cite{AhlswedeGacs}. 

\begin{definition}
\label{defn:q}
For $ \mathsf{p}\geq 1$ define
$$ \hspace*{1.5in} \mathsf{q}^*_{X;Y}(\mathsf{p}) := \inf\{\mathsf{q}:||\E[g(Y)|X]||_\mathsf{p}\leq ||g(Y)||_\mathsf{q}~ \forall g:\mathcal{Y}\mapsto \mathbb{R}\}. \hspace*{1.6in} \Box$$
\end{definition}
\begin{remark} Ahlswede and Gacs \cite{AhlswedeGacs} characterize hypercontractivity in terms of $s_\mathsf{p}(X,Y) := \frac{\mathsf{q}^*_{X;Y}(\mathsf{p})}{\mathsf{p}},$ for $p \geq 1$.
\end{remark}

If $r(x)$ and $p(x)$ are probability distributions on the same finite set,
we write $D \big(r(x) \| p(x) \big)$ for the relative entropy  distance of $r(x)$ 
from $p(x)$, i.e.
\[
D \big(r(x) \| p(x) \big) := \sum_x r(x) \log \frac{r(x)}{p(x)}~.
\]
To proceed to discuss the results of this paper, we need the following definition.
\begin{definition} Let $X$ and $Y$ be random variables with
 joint distribution $(X,Y) \sim p(x,y)$. We define
 \begin{equation}s^*(X;Y):=\sup_{r(x)\neq p(x)} \frac{D\big(r(y) \|
      p(y)\big)}{D\big(r(x) \| p(x)\big)},\label{eq:4}\end{equation}
where $r(y)$ denotes the $y$-marginal distribution of $r(x,y) :=r(x)p(y|x)$ and
the supremum on the right hand side is over all probability distributions $r(x)$ 
that are different from the probability distribution $p(x)$. If either $X$ or $Y$ is
a constant, we define $s^*(X;Y)$ to be $0$.
\hfill $\Box$
\end{definition}

{\it Remark}: From the data processing inequality for relative entropies it is immediate that $s^*(X;Y) \leq 1$. Further, $s^*(X;Y)$ can be regarded as a function of the input distribution $p(x)$ corresponding to a channel $p(y|x)$.

Below, we outline some of the properties of $ \mathsf{q}^*_{X;Y}(\mathsf{p})$ combining results from \cite{KA12} and from Theorems 3 and 5 in \cite{AhlswedeGacs}. 
\begin{theorem}
The following statements hold:
\begin{enumerate}[(a)]
\item For any fixed $\mathsf{p} > 1$, $ \mathsf{q}^*_{X;Y}(\mathsf{p}) \geq 1$ with equality {\it if and only if} $X$ and $Y$ are independent.
\item $ \mathsf{q}^*_{X;Y}(\mathsf{1}) = 1$ and $\frac{\mathsf{q}^*_{X;Y}(\mathsf{p})}{\mathsf{p}}$ is monotonically decreasing in $\mathsf{p}$.
\item $\frac{\mathsf{q}^*_{X;Y}(\mathsf{p})-1}{\mathsf{p}-1} \geq \rho_m^2(X;Y)$.
\item The chordal slope of  $\mathsf{q}^*_{X;Y}(\mathsf{p})$ at infinity, defined by $\lim_{\mathsf{p}\rightarrow\infty}\frac{\mathsf{q}^*_{X;Y}(\mathsf{p})-1}{\mathsf{p}-1}$, 
exists and is equal to  $s^*(X;Y)$. 
\item$\lim_{\mathsf{p}\downarrow 1}
  \frac{\mathsf{q}^*_{X;Y}(\mathsf{p})-1}{\mathsf{p}-1} = s^*(Y;X)$.
\end{enumerate}
\end{theorem}

\begin{remark}
Hypercontractive inequalities (and  their counterpart for $\mathsf{p} < 1$, called {\it reverse hypercontactive inequalities})  also play an important role in analysis, probability theory, and discrete Fourier analysis. Interested readers can refer to the introduction in \cite{Mossel11} for a brief summary of their development and impact in these areas. For results and applications of hypercontractivity and reverse hypercontractivity in information theory, interersted readers can refer to \cite{KA12}.
\end{remark}

\begin{figure}
\begin{center}
\includegraphics[width = 0.4\textwidth,height=!]{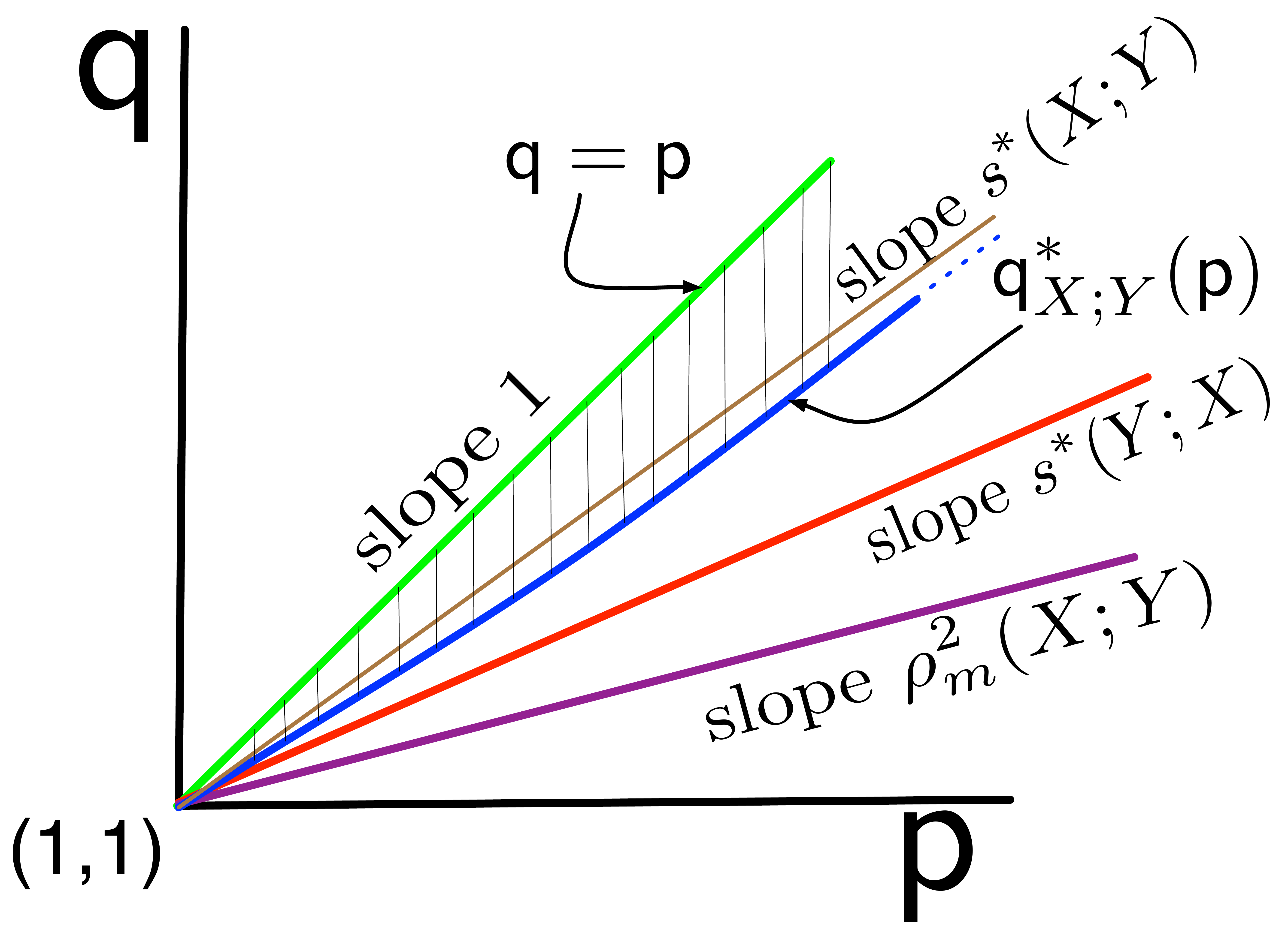}
\caption{The blue curve is an illustration of $q^*_{X;Y}(\mathsf{p})$
  (this curve is not convex in general).
The brown line represents the `chordal' slope
  $\frac{\mathsf{q}^*_{X;Y}(\mathsf{p})-1}{\mathsf{p}-1}$ as $p\to\infty,$ which turns out to be $s^*(X;Y).$ The red line is
  the slope of $q^*_{X;Y}(\mathsf{p})$ at $(1,1)$ defined by $\lim_{\mathsf{p} \downarrow 1} \frac{\mathsf{q}^*_{X;Y}(\mathsf{p})-1}{\mathsf{p}-1}$
 and turns out to be $s^*(Y;X)$. The
  purple line passes through $(1,1)$ and has slope $\rho_m^2(X;Y).$}
\label{fig:ribbon}
\end{center}
\end{figure}

In this paper we will provide alternate characterizations of both $\rho_m^2(X,Y)$ and $s^*(X;Y)$. Fix a channel $p(y|x),$ fix $\lambda \in [0,1]$, and consider the function\footnote{We abuse notation when we write $t_\lambda(X)$. We really
wish to think of $t_\lambda$ as a function of the probability distribution 
of $X$. }  of the probability distribution of $X$ denoted by $t_\lambda(X)$ 
which is defined by
$$ t_\lambda(X) := H(Y) - \lambda H(X).$$
 
We will show in Theorem \ref{Thm2} that
$\rho_m^2(X,Y)$ is the smallest $\lambda$ such that $t_\lambda(X)$ has a positive
semidefinite Hessian at $p(x)$ and $s^*(X;Y)$ is the smallest $\lambda$ such that $t_\lambda(X)$ matches its lower convex envelope, denoted by $\cxe[t_\la](X)$, at $p(x)$.

In \cite[Theorem 8]{ErkipCover} it was claimed that the following inequality holds:
$$ I(U;Y) \leq \rho_m^2(X;Y) I(U;X), ~\forall \ U-X-Y.$$
It turns out that this inequality is incorrect; we will provide a counter example in this paper. Further we will
show (Theorem \ref{Thm2}) that the following inequality holds, with a tight constant:
$$ I(U;Y) \leq s^*(X;Y) I(U;X), ~\forall \ U-X-Y.$$

The error in the proof in \cite[Theorem 8]{ErkipCover} seems to be a subtle, yet significant one. A similar error has also occurred in \cite{HuangZheng}, where the authors independently rediscover the erroneous result of \cite[Theorem 8]{ErkipCover} using similar techniques.

\subsection{Alternate characterizations of the Hirschfeld-Gebelein-R\'{e}nyi maximal correlation}

In this section we will review some alternate characterizations of the Hirschfeld-Gebelein-R\'{e}nyi maximal correlation which are known in the literature.

\subsubsection{R\'{e}nyi's characterization}:
As mentioned earlier, R\'{e}nyi derived the following ``one-function'' alternate characterization for $\rho_m(X;Y)$ \cite{Renyi}:
\begin{equation}
\rho_m^2(X;Y)=\max_{f(X): \E f(X)=0,
    \E[f^2(X)]=1}\E[\E[f(X)|Y]^2].\label{eq:2}\end{equation}
The validity of this characterization can be proved by fixing $f$ with 
$\E[f(X)] = 0$ and $\E[f^2(X)] = 1$ and
showing that setting $g(Y) = \alpha \E[f(X)|Y]$ maximizes 
$\E[f(X)g(Y)]$ among all functions $g$ with $\E[g(Y)] = 0$ and
$\E[g^2(Y)] =1$ when $\alpha \geq 1$ is chosen so that $\alpha^2 \E(\E[f(X)|Y]^2) = 1$. 
This is a simple consequence of the Cauchy-Schwartz inequality.

\subsubsection{Distribution simulation characterization}
Consider a random variable $X'$ such that $X - Y - X'$ is Markov and $(X,Y) \stackrel{d}{=} (X^\prime,Y).$
Then
\begin{equation}\rho_m^2(X;Y)=\max_{f(X): \E f(X)=0,
    \E[f^2(X)]=1}\E[f(X)f(X^\prime)].\label{eq:dist}\end{equation}
This result follows from R\'{e}nyi's characterization which was given in 
  \eqref{eq:2}
above. 
Since $(X,Y) \stackrel{d}{=} (X^\prime,Y),$ we have
$\E[f(X)|Y] = \E[f(X^\prime)|Y].$ Hence $\E[\E[f(X)|Y]^2] = \E[\E[f(X)|Y]\E[f(X^\prime)|Y]] \stackrel{(a)}{=}
\E[\E[f(X)f(X^\prime)|Y]] = \E[f(X)f(X^\prime)],$ where (a) holds
because $X-Y-X^\prime.$

\subsubsection{Singular value characterization}
For finite valued random variables maximal correlation $\rho_m(X;Y)$ can also be characterized \cite{Witsenhausen75} by the second largest
singular value of the matrix $Q$ with entries 
$Q_{x,y} = \frac{p(x,y)}{\sqrt{p(x)p(y)}}$.  
This result can be seen by writing $\E[f(X)g(Y)]$ as 
$\sum_{x,y} (f(x) \sqrt{p(x)}) Q(x,y) (g(y) \sqrt{p(y)})$, observing that 
$\sum_x \sqrt{p(x)} Q(x,y) = \sqrt{p(y)}$ and $\sum_y Q(x,y) \sqrt{p(y)} = \sqrt{p(x)}$,
and that the conditions $\E[f(X)] = 0$ and $\E[g(Y)] = 0$ are respectively equivalent to 
requiring that $x \mapsto f(x) \sqrt{p(x)}$ is orthogonal to $x \mapsto \sqrt{p(x)}$
and that $y \mapsto g(y) \sqrt{ p(y)}$ is orthogonal to $y \mapsto \sqrt{p(y)}$.\\

There is a simple formula for $\rho_m(X;Y)$ if at least one of $X$ or $Y$ is binary-valued, 
which is most easily seen by using the singular value characterization:
\begin{equation}\rho_m^2(X;Y) =\left[\sum_{x,y} \frac{p(x,y)^2}{p(x)p(y)} \right] -
1.\label{eq:eigenvalue}\end{equation}
This follows from observing that $\rho_m^2(X;Y)$ is the
second largest eigenvalue of both $QQ^{T}$ and
$Q^{T}Q.$ If one of these is a 2 by 2 matrix, we can find the
second largest eigenvalue by computing the trace and subtracting the largest eigenvalue, i.e. 1, from it. 

\subsection{Properties of $\rho_m(X;Y)$}
In this section, we will present some known properties of the maximal correlation $\rho_m(X;Y)$.

\subsubsection{Tensorization of $\rho_m(X;Y)$}

The following theorem shows that maximal correlation
\emph{tensorizes}. It was proved by Witsenhausen in
\cite{Witsenhausen75}.  For a function of probability distributions to have the property of the first
sentence of the theorem is what it means to say that it tensorizes.
\begin{theorem}\label{thm:mc-tensorization} (Witsenhausen \cite{Witsenhausen75})
If $(X_1,Y_1), (X_2,Y_2)$ are independent, then
$$\rho_m(X_1,X_2;Y_1,Y_2)=\max\{\rho_m(X_1;Y_1), \rho_m(X_2;Y_2)\}.$$
In particular if $(X_1,Y_1), (X_2,Y_2)$ are i.i.d., then $\rho_m(X_1,X_2;Y_1,Y_2)=\rho_m(X_1;Y_1).$
\hfill $\Box$
\end{theorem}

The elegant proof in \cite{Kumar10} (for finite valued random variables) uses the singular value characterization and is reproduced below. When $(X_1,Y_1)$
is independent of $(X_2,Y_2)$ it is immediate that the matrix $Q$
defined by $Q_{x_1,x_2,y_1,y_2} =
\frac{p_1(x_1,y_1)p_2(x_2,y_2)}{\sqrt{p_1(x1)p_2(x_2)p_1(y_1)p_2(y_2)}}$
is the Kronecker product of the corresponding individual matrices
$\hat{Q}_{x_1,y_1} = \frac{p_1(x_1,y_1)}{\sqrt{p_1(x_1)p_1(y_1)}}$ and
$\tilde{Q}_{x_2,y_2} = \frac{p_2(x_2,y_2)}{\sqrt{p_2(x_2)p_2(y_2)}}$,
i.e. $Q = \hat{Q} \otimes \tilde{Q}$. It is known that the singular
values of $Q$ are given as the set of products of one singular value of $\hat{Q}$
with one singular value of $\tilde{Q}$. Since the largest singular values of each of the three
matrices is unity, it is immediate that the second largest singular
value of $Q$ is $\max\{\rho_m(X_1;Y_1), \rho_m(X_2;Y_2)\}.$

Witsenhausen
\cite{Witsenhausen75} showed that the maximal correlation of two
random variables gives the answer to the following problem:
consider two agents, the first of whom observes $X^n$, while the second
observes $Y^n$, where $(X_i, Y_i), 1 \le i \le n,$ are
i.i.d. copies of $(X,Y)$. Each agent makes a
binary decision based on the sequence available to it. The entropy
of the each binary decision should be bounded away from zero by a
constant. Witsenhausen showed that the probability of agreement between these
decisions can be made to
converge to $1$, as $n$ converges to infinity, if and only if
$\rho_m(X;Y)=1$. This is a version of the main result in the path-breaking work of
G\'{a}cs and K\"{o}rner \cite{GK}, which introduced the concept of 
G\'{a}cs-K\"{o}rner common information.

Erkip and Cover \cite{ErkipCover} studied the problem of investment in the 
stock market with side information of limited rate with the aim of quantifying the 
value of the side information in improving the growth rate of wealth. In one part of their
much broader contribution, they present a data processing inequality which claims that
$$ I(U;Y) \leq \rho_m^2(X;Y) I(U;X), ~\forall \ U-X-Y$$
where $\rho_m(X;Y)$ is the Hirschfeld-Gebelein-R\'{e}nyi maximal correlation 
between the random
variables $X$ and $Y$. As we stated earlier, this inequality is incorrect.

Kang and Ulukus illustrated some applications of
maximal correlation in distributed source and channel coding problems
\cite{KangUlukus}.  Beigi has introduced a quantum version of the
maximal correlation for bipartite quantum states, and has shown that
this measure fully characterizes bipartite states from which common
randomness distillation under local operations is possible
\cite{Beigi}. 

Recently Kamath and Anantharam \cite{KA12} have used maximal
correlation to study the problem of non-interactive simulation of
joint distributions.  They also used hypercontractivity and reverse
hypercontractivity to show that under certain conditions these can
provide stronger impossibility results for the simulation problem than
those obtained by maximal correlation.

\subsection{Alternate characterization and properties of $\mathsf{q}^*_{X;Y}(\mathsf{p})$}

In \cite{KA12}, the authors defined the following region which can be used to characterize $\mathsf{q}^*_{X;Y}(\mathsf{p})$.
\begin{definition} 	\label{defn:3}
For a pair of random variables $(X,Y)\sim p(x,y)$ on
  $\mathcal{X}\times\mathcal{Y},$  the \emph{hypercontractivity ribbon} is the subset
  $$\mathcal{R}_{X;Y}\subseteq \{(\mathsf{p},\mathsf{q})\in\mathbb{R}^2: 1\leq \mathsf{q}\leq \mathsf{p}\mbox{ or }1\geq
  \mathsf{q}\geq \mathsf{p}\}$$ 
defined by\footnote{This characterization of the hypercontractivity ribbon is given in \cite{KA12}. Another characterization, which is closer to how hypercontractivity is normally
discussed in the literature, will be mentioned later.}
  \begin{itemize}
  \item$(1,1)\in\mathcal{R}_{X;Y};$
  \item For $1\leq \mathsf{q} \le \mathsf{p},$ $(\mathsf{p},\mathsf{q})\in \mathcal{R}_{X;Y}$ iff 
    \begin{align}\label{eq:hc2}\E f(X)g(Y)\leq
      ||f(X)||_{\mathsf{p}^\prime}||g(Y)||_\mathsf{q}&\qquad \forall
      f:\mathcal{X}\mapsto \mathbb{R}, g:\mathcal{Y}\mapsto \mathbb{R};~~\quad\end{align}
  \item For $1\geq \mathsf{q} \ge \mathsf{p},$  $(\mathsf{p},\mathsf{q})\in \mathcal{R}_{X;Y}$ iff 
    \begin{align}\label{eq:rhc2}\E f(X)g(Y)\geq
      ||f(X)||_{\mathsf{p}^\prime}||g(Y)||_\mathsf{q}&\qquad\forall
      f:\mathcal{X}\mapsto (0,\infty), g:\mathcal{Y}\mapsto (0,\infty).\end{align}
  \end{itemize}

  When $1\leq \mathsf{q} < \mathsf{p},$ inequalities such as
  \eqref{eq:hc2} are referred to in the literature as {\it
    hypercontractive} inequalities and when $1\geq \mathsf{q}>
  \mathsf{p},$ inequalities such as \eqref{eq:rhc2} are referred to as
  {\it reverse hypercontractive} inequalities.
\hfill $\Box$
\end{definition}

Then one can alternatively define $\mathsf{q}^*_{X;Y}(\mathsf{p})$ according to
$$ \mathsf{q}^*_{X;Y}(\mathsf{p}) := \inf\{\mathsf{q}: (\mathsf{q},\mathsf{p}) \in \mathcal{R}_{X;Y}\}, \mathsf{p} \geq 1.$$

The equivalence of this characterization to that in definition \ref{defn:q} is 
proved in \cite{KA12}. The proof is similar to that of Renyi's alternate characterization of $\rho_m(X,Y)$ and is a straightforward application of H\"{o}lder's inequality. 


  Likewise, we can define $\mathcal{R}_{Y;X}.$ In general,
  $\mathcal{R}_{X;Y} \neq \mathcal{R}_{Y;X},$ but the two are related
  by an intimate duality relationship that is clear from
  \eqref{eq:hc2} and \eqref{eq:rhc2}:
  $$(\mathsf{p},\mathsf{q})\in\mathcal{R}_{X;Y} \iff
  (\mathsf{q}^\prime,\mathsf{p}^\prime)\in\mathcal{R}_{Y;X}.$$ Using this duality relationship \cite{KA12} establishes that $\lim_{\mathsf{p}\to 1}
  \frac{\mathsf{q}^*_{X;Y}(\mathsf{p})-1}{\mathsf{p}-1} =
  \left.\frac{d}{d\mathsf{p}}\mathsf{q}^*_{X;Y}(\mathsf{p})\right|_{\mathsf{p}=1}
  = s^*(Y;X)$.
\begin{remark}\label{remark:unequal}
  In general, $s^*(X;Y)\neq s^*(Y;X)$ as shown by the following
  example.  Let $(X,Y)$ be 0-1 valued with $\mathbb{P}(X=0) = 0.85,
  \mathbb{P}(Y=0) = 0.39, \mathbb{P}(X=Y=0) = 0.36.$ Then, computation
  gives us $s^*(X;Y) = 0.045..., s^*(Y;X) = 0.029...$.
\end{remark}

Most of the applications of hypercontractivity traces its roots to the following {\it tensorization} property of the hypercontractive ribbon.
\begin{theorem}\label{thm:hc-tensorization} 
  (\cite{AhlswedeGacs, Mossel11}) 
If $(X_1,Y_1)$ and $(X_2,Y_2)$ are independent, then  $\mathcal{R}_{(X_1,X_2);(Y_1,Y_2)}=\mathcal{R}_{X_1;Y_1}\cap
  \mathcal{R}_{X_2;Y_2}.$ In particular, if $(X_1,Y_1)$ and $(X_2,Y_2)$ are i.i.d., then
  $\mathcal{R}_{(X_1,X_2);(Y_1,Y_2)}=\mathcal{R}_{X_1;Y_1}.$
\hfill $\Box$
\end{theorem}

Theorem \ref{thm:hc-tensorization} can be thought of as saying that the
whole hypercontractivity ribbon tensorizes, 
since it says that for each $(\mathsf{p},\mathsf{q})$ we have 
\[
\mathds{1}((\mathsf{p},\mathsf{q}) \notin \mathcal{R}_{(X_1,X_2);(Y_1,Y_2)}) = 
\max\{ \mathds{1}((\mathsf{p},\mathsf{q}) \notin \mathcal{R}_{X_1;Y_1}), 
\mathds{1}((\mathsf{p},\mathsf{q}) \notin \mathcal{R}_{X_2;Y_2})\}~.
\]
A consequence of this then is that $s^*(X;Y)$ tensorizes, i.e. for
$(X_1,Y_1)$ and $(X_2,Y_2)$ independent,
\[
s^*(X_1,X_2;Y_1,Y_2) = \max\{s^*(X_1;Y_1), s^*(X_2;Y_2)\}.
\]
We will give a  alternate proof of the tensorization of $s^*(X;Y)$ later using our new characterization involving the function $t_\lambda(X)$ that was introduced earlier. A direct proof of this tensorization can be obtained as follows. The direction $s^*(X_1,X_2;Y_1,Y_2) \geq \max\{s^*(X_1;Y_1), s^*(X_2;Y_2)\}$ is immediate; hence we only show the non-trivial direction. Note that for any $r(x_1,x_2) \neq p(x_1,x_2)$ we have
\begin{align*}
\frac{D(r(y_1,y_2)|| p(y_1,y_2))}{ D(r(x_1,x_2)|| p(x_1,x_2))} & \stackrel{(a)}{=} \frac{ D(r(y_1)|| p(y_1)) + \sum_{y_1} r(y_1) D(r(y_2|y_1)||p(y_2))  }{   D(r(x_1)|| p(x_1)) + \sum_{x_1} r(x_1) D(r(x_2|x_1)||p(x_2))  }\\
& = \frac{ D(r(y_1)|| p(y_1)) + \sum_{y_1} r(y_1) D\left(\sum_{x_1}r(x_1|y_1)r(y_2|x_1)||p(y_2)\right)  }{   D(r(x_1)|| p(x_1)) + \sum_{x_1} r(x_1) D(r(x_2|x_1)||p(x_2))  }\\
&\stackrel{(b)}{\leq}  \frac{ D(r(y_1)|| p(y_1)) + \sum_{y_1} \sum_{x_1} r(x_1|y_1)r(y_1) D\left(r(y_2|x_1)||p(y_2)\right)  }{   D(r(x_1)|| p(x_1)) + \sum_{x_1} r(x_1) D(r(x_2|x_1)||p(x_2))  }\\
&=  \frac{ D(r(y_1)|| p(y_1)) + \sum_{x_1} r(x_1)  D\left(r(y_2|x_1)||p(y_2)\right)  }{   D(r(x_1)|| p(x_1)) + \sum_{x_1} r(x_1) D(r(x_2|x_1)||p(x_2))  }\\\\
& \leq \max\{s^*(X_1;Y_1), s^*(X_2;Y_2)\}.
\end{align*}
In the above $(a)$ uses the fact that $p(x_1y_1,x_2,y_1)=p_1(x_1,y_1)p_2(x_2,y_2)$ and $(b)$ uses the convexity of $D(p||q)$ in $p$. The last inequality follows from the definition of $s^*(X_1;Y_1), s^*(X_2;Y_2),$ and our assumption that $r(x_1,x_2) \neq p(x_1,x_2)$ which guarantees that at least one of the terms in the denominator is non-zero. Finally taking $\sup$ over all such  $r(x_1,x_2)$ we obtain the non-trivial direction $s^*(X_1,X_2;Y_1,Y_2) \leq \max\{s^*(X_1;Y_1), s^*(X_2;Y_2)\}.$

\section{Main Results}
\label{mainresult}
One of the main contributions of this paper is a correction to the data processing
inequality claimed by Erkip and Cover in \cite[Theorem 8]{ErkipCover}.
We provide a counterexample to their claim
and point out a location in their proof where the argument is incomplete.
We then find the correct constant to get a tight data processing inequality of the 
type they considered.

\subsection{Counterexample to the Erkip-Cover data processing inequality}
In \cite[Theorem 8]{ErkipCover}, Erkip and Cover claimed that 
\begin{equation}
I(U;Y)\leq \rho_m^2(X;Y) I(U;X)
\label{eq:erkip}
\end{equation} holds whenever $U-X-Y$ form a Markov chain. Furthermore they claimed that, $\rho_m^2(X;Y)$ is the minimum such constant, i.e. 
\begin{equation} \sup_{U:~U-X-Y, I(U;X) > 0}\frac{I(U;Y)}{I(U;X)}=\rho_m^2(X;Y).\label{eq:erkipmin}\end{equation} 

We will first provide a counterexample to these claims
and then point where there is a gap in their argument.
 In a subsequent subsection we will identify $s^*(X;Y)$ as the correct constant to replace $\rho_m^2(X;Y)$ in \eqref{eq:erkip} and \eqref{eq:erkipmin}.

\subsubsection{Counterexample to \eqref{eq:erkip} and \eqref{eq:erkipmin}}
\begin{figure}
\begin{center}
\begin{tikzpicture}
\filldraw (1,1) circle (3pt);
\filldraw(1,-1) circle (3pt);
\filldraw (4,2) circle (3pt);
\filldraw(4,-2) circle (3pt);
\filldraw(4,0) circle (3pt);
\node at (0.5,1) {$0$};
\node at (0.5,-1) {$1$};
\node at (4.5,2) {$0$};
\node at (4.5,-2) {$1$};
\node at (2.5,1.9) {$\frac 23$};
\node at (2.5,0.8) {$\frac 13$};
\node at (2.5,-0.8) {$\frac 12$};
\node at (2.5,-1.9) {$\frac 12$};
\node at (0.5,-3) {$X$};
\node at (4.5,-3) {$Y$};
\node at (4.5,0) {$E$};
\draw [->,thick] (1,1) -- (3.9,2);
\draw [->,thick] (1,1) -- (3.9,0);
\draw [->,thick] (1,-1) -- (3.9,-2);
\draw [->,thick] (1,-1) -- (3.9,0);
\end{tikzpicture}
\caption{An asymmetric erasure channel.}
\label{figC}
\end{center}
\end{figure}
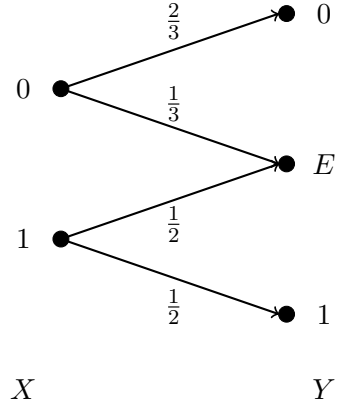

 Let $X$ be a binary random variable with $p(X=0) = \frac 12$. Define $p(x,y)$ by passing $X$ through the asymmetric erasure channel given in Fig. \ref{figC}.
Using Equation \eqref{eq:eigenvalue}, one can verify for this pair $(X,Y)$ 
that $\rho_m^2(X;Y) = 0.6$. 

Suppose we construct
$U$ satisfying $U-X-Y$ such that $U|X=0\sim\Ber(0.1),
U|X=1\sim\Ber(0.4).$ Then $I(U;Y) = 0.055770...$ and $I(U;X) = 0.09130...,$
so that $\frac{I(Y;U)}{I(X;U)}=0.6108... > 0.6 = \rho_m^2(X;Y),$ and
this contradicts  \eqref{eq:erkipmin}.

It can be shown in a reasonably straightforward manner, using our
characterization in Theorem \ref{Thm2}, that $s^*(X;Y) =
\frac{1}{2}\log_2 \left(\frac{12}{5}\right) = 0.631517...$ for this
pair of random variables $(X,Y)$. Simulation shows that for a suitable
sequence of $U_i$ with $I(U_i;X)\to 0,$ we can have
$\frac{I(U_i;Y)}{I(U_i;X)}$ approach $s^*(X;Y)$ for this example. A
sequence of such $U_i$ is shown in the table below.

\begin{center}
\begin{tabular}{|c|c|c|c|c|}
\hline 
$P(U=1|X=0)$ & $P(U=1|X=1)$ & $I(U;Y)$ & $I(U;X)$ & $\frac{I(U;Y)}{I(U;X)}$\\ \hline
0.1 & 0.4 & 0.055770... & 0.09130... & 0.6108... \\ \hline
0.01 & 0.23 & 0.062321... & 0.099958... & 0.6234... \\ \hline
0.001 & 0.102 & 0.031038... & 0.049379... & 0.6285... \\ \hline
0.0001 & 0.04 & 0.012507... & 0.019838... & 0.6304... \\ \hline
0.00001 & 0.01474 & 0.0046418... & 0.0073545... & 0.6311... \\ \hline
0.000001 & 0.005232 & 0.0016507... & 0.0026145... & 0.6313... \\\hline
0.0000001 & 0.0018146 & 0.00057285... & 0.00090716... & 0.6314... \\\hline
0.00000001 & 0.00061973 & 0.000195672... & 0.000309852... & 0.63150... \\\hline
\end{tabular}
\end{center}

The error of the Erkip-Cover proof seems to lie in their use of a 
Taylor's series expansion. Consider the expansion in the left column of page 1037 
of their paper \cite{ErkipCover}, where
they use their equation (16) to expand around $p(\tilde v)$. It is possible that $p(\tilde v)$ is zero for some $\tilde v$ and this causes an error as the 
derivative in this direction is infinity and the Taylor's series expansion is no longer valid. As our counterexample shows  this seems to be a significant but subtle error that cannot be worked around. 

Some of the works that use this incorrect result of \cite{ErkipCover},
such as \cite{Courtade} and \cite{ZhaoChia}, are affected by this error. A
claim similar to that of \cite{ErkipCover}, which appears in
\cite{HuangZheng}, is also false.\footnote{This paper studies the
  ratio $\frac{I(U;Y)}{I(U;X)}$ when $I(U;X)$ is very small. However,
  as pointed out in \cite{ErkipCover}, the supremum of
  $\frac{I(U;Y)}{I(U;X)}$ occurs when $I(U;X)\rightarrow 0$. So the
  problem studied by \cite{HuangZheng} is the same as that of
  \cite{ErkipCover}.}

\subsection{A geometric characterization of $\rho_m^2(X;Y)$ and $s^*(X;Y)$}
Given $p(x,y)$, we can 
treat $p(y|x)$ 
as a channel,  and then consider the function of the input distribution $p(x)$, 
defined by
$$ t_{\lambda}(X) := H(Y) - \lambda H(X),$$
where $\lambda$ is a constant in $[0,1]$. Observe that the function is concave when $\lambda=0$ and convex when $\lambda=1$.\footnote{This convexity at $\lambda = 1$ follows from the fact that for any $U-X-Y$ we have $I(U;X)\geq I(U;Y)$ or equivalently $H(Y)-H(X)\leq H(Y|U)-H(X|U)$.} 

We write $\cxe[t_\la](X)$ for the convex hull of $t_\la(X)$.
If $\cxe[t_\la](X) = t_\la(X)$ at $p(x)$ for some $\lambda$, then note that for any $\la_1 \geq \la$
\begin{align*} 
\cxe[t_{\la_1}](X)
&= \cxe[t_{\la} - (\la_1 - \la) H](X)\\
& \geq 
\cxe[t_\la](X)
- (\la_1 - \la) H(X). 
\end{align*}
Here the inequality comes from  $\cxe[f+g] \geq \cxe[f] + \cxe[g]$ and since  $- (\la_1 - \la) H(X)$ is convex.
Therefore at $p(x)$ we will have that 
$$ t_{\la_1}(X) \geq \cxe[t_{\la_1}](X) \geq \cxe[t_\la](X)  - (\la_1 - \la) H(X) = t_\la(X) - (\la_1 - \la) H(X) = t_{\la_1}(X).$$
Thus we see that if  $\cxe[t_\la](X) = t_\la(X)$ at $p(x)$ for some $\lambda$ then $\cxe[t_{\la_1}](X) = t_{\la_1}(X)$ at $p(x)$ for all $\la_1 \geq \la$.

 The following theorem gives a geometric interpretation of $\rho_m^2(X;Y)$ and $s^*(X;Y)$ in terms of the behaviour of the function $t_\la(X)$ and identifies $s^*(X;Y)$ as the correct replacement for $\rho^2_m(X;Y)$ in \eqref{eq:erkip} and \eqref{eq:erkipmin}.

\begin{theorem}\label{Thm2} The following statements hold:
\begin{enumerate}
\item $\rho_m^2(X;Y)$ is the minimum value of $\lambda$ such that the function $t_\la(X)$ 
has a positive semidefinite Hessian at $p(x)$. 
\item $s^*(X;Y)$ is the minimum value of $\lambda$ such that the function $t_\la(X)$ touches its lower convex envelope at $p(x)$, i.e. such that
$\cxe[t_\la](X) = t_\la(X)$ at $p(x)$. Furthermore, 
$$\sup_{U:~U-X-Y, I(U;X) > 0}\frac{I(U;Y)}{I(U;X)} = s^*(X;Y).$$
\end{enumerate}
\hfill $\Box$
\end{theorem}

\begin{IEEEproof}[Proof of 1)] This follows from R\'{e}nyi's characterization of the 
maximal correlation, given in \eqref{eq:2} above.
Take an arbitrary multiplicative perturbation of the form
$p_{\epsilon}(x)=p(x)(1+\epsilon f(x))$. For $p_{\epsilon}$ to stay a
valid perturbation we need $\E[f(X)]=0$. Furthermore we can normalize
$f$ by assuming that $\E[f^2(X)]=1$. The second derivative 
in $\epsilon$ of $H(Y)-\lambda H(X)$ is equal to \cite{GA12}
$$-\E[\E[f(X)|Y]^2]+\lambda \E[f^2(X)]=-\E[\E[f(X)|Y]^2]+\lambda~,$$
which is non-negative as long as $\lambda \geq \E[\E[f(X)|Y]^2]$. Thus  the minimum value $\lambda^*$ such that the second derivative is non-negative for all local perturbations is 
$$\lambda^* = \max_{f(X): \E f(X)=0,
    \E[f^2(X)]=1}\E[\E[f(X)|Y]^2] = \rho^2_m(X;Y),$$
where the last equality follows from R\'{e}nyi's characterization of maximal correlation.
\end{IEEEproof}

\begin{IEEEproof}[Proof of 2)]
Consider the minimum value of $\lambda$, say $\lambda^\dagger$, such that the function $t_\la(X)$ touches its lower convex envelope at $p(x)$. Thus, 
equivalently we are looking for the minimum $\lambda$ such that for $(X,Y)\sim p(x,y)$ 
we have
$$ H(Y)-\lambda H(X)\leq H(Y|U)-\lambda H(X|U),\qquad \forall \ U: U-X-Y.$$

Note that if $U$ is independent of $X$, i.e. $I(U;X) = 0$ then the above inequality is always true. Equivalently we require the minimum $\lambda$ such that,
$$\lambda\geq \frac{I(U;Y)}{I(U;X)}, \qquad  \forall \ U: U-X-Y~\mbox{with}~ I(U;X) > 0.$$
Thus, 
$$\lambda^\dagger = \sup_{U:~U-X-Y, I(U;X) > 0}\frac{I(U;Y)}{I(U;X)}.$$

{\it Remark}: Since $t_\lambda(X) = \cxe[t_\la](X)$ at $p(x)$ implies
that the Hessian of $t_\la(X)$ at $p(x)$ is positive semidefinite,
we have
\begin{equation}\sup_{U:~U-X-Y, I(U;X) > 0}\frac{I(U;Y)}{I(U;X)}\geq \rho_m^2(X;Y).\label{eq:inq1}\end{equation}

It remains to show that $\lambda^\dagger = s^*(X;Y)$ or equivalently that
$$\sup_{U:~U-X-Y, I(U;X) > 0}\frac{I(U;Y)}{I(U;X)}=s^*(X;Y).$$

From standard cardinality bounding arguments, it suffices to consider $|\mathcal{U}| \leq |\mathcal{X}|+1$ to determine the value of  $\sup_{U:~U-X-Y, I(U;Y) > 0}\frac{I(U;Y)}{I(U;X)}$.\footnote{Indeed our proof below indicates that even a binary $U$ suffices.} 

For any $|\mathcal{U}| \leq |\mathcal{X}| + 1$ and $U-X-Y$ a Markov chain
with $I(U;X) > 0$  and $X \sim p(x)$, denote $\mathrm{P}(U=u) =: w_u, \mathrm{P}(X=x|U=u) =: r_u(x)$. Clearly $\sum_u w_u r_u(x) = p(x)$. Let the channel-induced distributions on $Y$ corresponding to the $r_u(x)$ be denoted by $r_u(y)$ respectively. 
Then elementary manipulations yield 
\begin{equation*}
  \frac{I(U;Y)}{I(U;X)} =  \frac{\sum_{u\in\mathcal{U}~:~r_u(x) \neq p(x)} w_u
    D(r_u(y)||p(y))}{\sum_{u\in\mathcal{U}~:~r_u(x) \neq p(x)} w_u D(r_u(x)||p(x))} \leq \sup_{r(x) \neq p(x)} \frac{D\big(r(y) \| p(y)\big)}{D\big(r(x) \| p(x)\big)}~,
\end{equation*}
where $r(y)$ denotes the channel-induced probability distribution on $Y$ corresponding
to the probability distribution $r(x)$ on $X$.

Since the above holds for all $U$ such that $U-X-Y$ is a Markov chain and $I(U;X) > 0$, 
we have
$$\sup_{U:~U-X-Y, I(U;X) > 0}\frac{I(U;Y)}{I(U;X)} \leq \sup_{r(x)} \frac{D\big(r(y) \| p(y)\big)}{D\big(r(x) \| p(x)\big)} = s^*(X;Y),$$
where the last equality is by definition, see \eqref{eq:4} above.

To show the other direction, we assume that $s^*(X;Y) > 0$, else there is nothing to prove. 
Let $\delta \in (0,s^*(X;Y))$ be arbitrary. 
We also assume without loss of generality that $p(x) > 0 ~\forall x \in \mathcal{X}$ and 
$p(y) > 0 ~\forall y \in \mathcal{Y}$, since otherwise
we could have simply changed the definition of $\mathcal{X}$ and
$\mathcal{Y}$.

Let $\mathcal{U}_\epsilon := \{1,2\}$.
Fix a sufficiently small
  $\epsilon >0$ and define $U_\epsilon$ by:
  \begin{itemize}
  \item $w_1 = \epsilon, r_1(x) = r^*(x),$
  \item $w_2 = 1-\epsilon, r_2(x) = p(x) + \frac{\epsilon}{1-\epsilon}(p(x)-r^*(x))= \frac{1}{1-\epsilon}p(x) - \frac{\epsilon}{1-\epsilon}r^*(x)$,
  \end{itemize}
where $r^*(x) \neq p(x)$ is a probability distribution satisfying
$ \frac{D\big(r^*(y) \| p(y)\big)}{D\big(r^*(x) \| p(x)\big)} > s^*(X;Y) - \delta .$
  For sufficiently small $\epsilon > 0,$ we will have that $r_2(x)$ is
  a probability distribution. 
     Note  that $w_1+w_2=1$ and $w_1r_1(x) +
  w_2r_2(x) = p(x)\ \forall x\in\mathcal{X}.$ Clearly $I(U_\epsilon;Y) > 0$,
since $I(X;Y) = 0$ would have implied that $s^*(X;Y) = 0$.

For any $0 < \lambda < s^*(X;Y)-\delta$ define the function
$$ g(\epsilon) := I(U_\epsilon;Y) - \lambda I(U_\epsilon;X).$$

We have
\begin{align*}
\frac{dg(\epsilon)}{d\epsilon} &= - \frac{d}{d\epsilon} \left( \epsilon H(r^*(y)) + (1-\epsilon) H\left(\frac{1}{1-\epsilon}p(y) - \frac{\epsilon}{1-\epsilon}r^*(y)\right)  \right) \\
& \quad + \lambda \frac{d}{d\epsilon} \left( \epsilon H(r^*(x)) + (1-\epsilon) H\left(\frac{1}{1-\epsilon}p(x) - \frac{\epsilon}{1-\epsilon}r^*(x)\right)  \right) \\
& = - H(r^*(y)) + H\left( \frac{p(y) - \epsilon r^*(y)}{1-\epsilon}\right) + \lambda H(r^*(x)) - \lambda H\left( \frac{p(x) - \epsilon r^*(x)}{1-\epsilon}\right) \\
& \quad - \sum_y \frac{r^*(y)-p(y)}{1-\epsilon} \log \left( \frac{p(y) - \epsilon r^*(y)}{1-\epsilon}\right)  + \lambda \sum_x  \frac{r^*(x)-p(x)}{1-\epsilon} \log \left( \frac{p(x) - \epsilon r^*(x)}{1-\epsilon}\right) .
\end{align*}

Thus 
$$ \frac{dg(\epsilon)}{d\epsilon}\Big|_{\epsilon=0} =  D\big(r^*(y) \| p(y)\big) - \lambda D\big(r^*(x) \| p(x)\big)   > 0,  $$
where the last inequality is because
$0 < \lambda < s^*(X;Y) - \delta$ and $ \frac{D\big(r^*(y) \| p(y)\big)}{D\big(r^*(x) \| p(x)\big)} > s^*(X;Y) - \delta .$ Since $g(0) =0$ this implies that for some $\epsilon' > 0$ we have $I(U_{\epsilon'};Y) - \lambda I(U_{\epsilon'};X) > 0$ or that
$$ \sup_{U:~U-X-Y, I(U;Y) > 0}\frac{I(U;Y)}{I(U;X)}  \geq \frac{I(U_{\epsilon'};Y)}{I(U_{\epsilon'};X)} > \lambda.   $$

Since the above holds for all $\lambda < s^*(X;Y) - \delta$ we have $$ \sup_{U:~U-X-Y, I(U;Y) > 0}\frac{I(U;Y)}{I(U;X)} \geq s^*(X;Y)-\delta.$$
Finally, since $\delta > 0$ is arbitrary, we are done.
\end{IEEEproof}

\medskip

{\it Remarks}:
\begin{itemize}
\item Note that $\rho_m^2(X;Y)$ is symmetric in the pair $(X,Y)$ but
  $s^*(X;Y)$ is not, i.e. $s^*(X;Y) \neq s^*(Y;X)$ in general. Thus,
  $\sup_{U:~U-X-Y, I(U;X) > 0}\frac{I(U;Y)}{I(U;X)} \neq
  \sup_{V:~X-Y-V, I(V;Y) > 0}\frac{I(V;X)}{I(V;Y)}$ in general, which is a qualitatively
  different phenomenon than predicted by the incorrect Erkip-Cover
  claim in  \eqref{eq:erkipmin} above.
\item This theorem also explains the motivation for our counterexample of the previous subsection. The plot of $p(x)\mapsto H(Y) - 0.6 H(X)$ for the channel $p(y|x)$ described earlier is given in Fig. \ref{fig1}.
\begin{figure}\label{fig1}
\begin{center}
\includegraphics[scale=0.4]{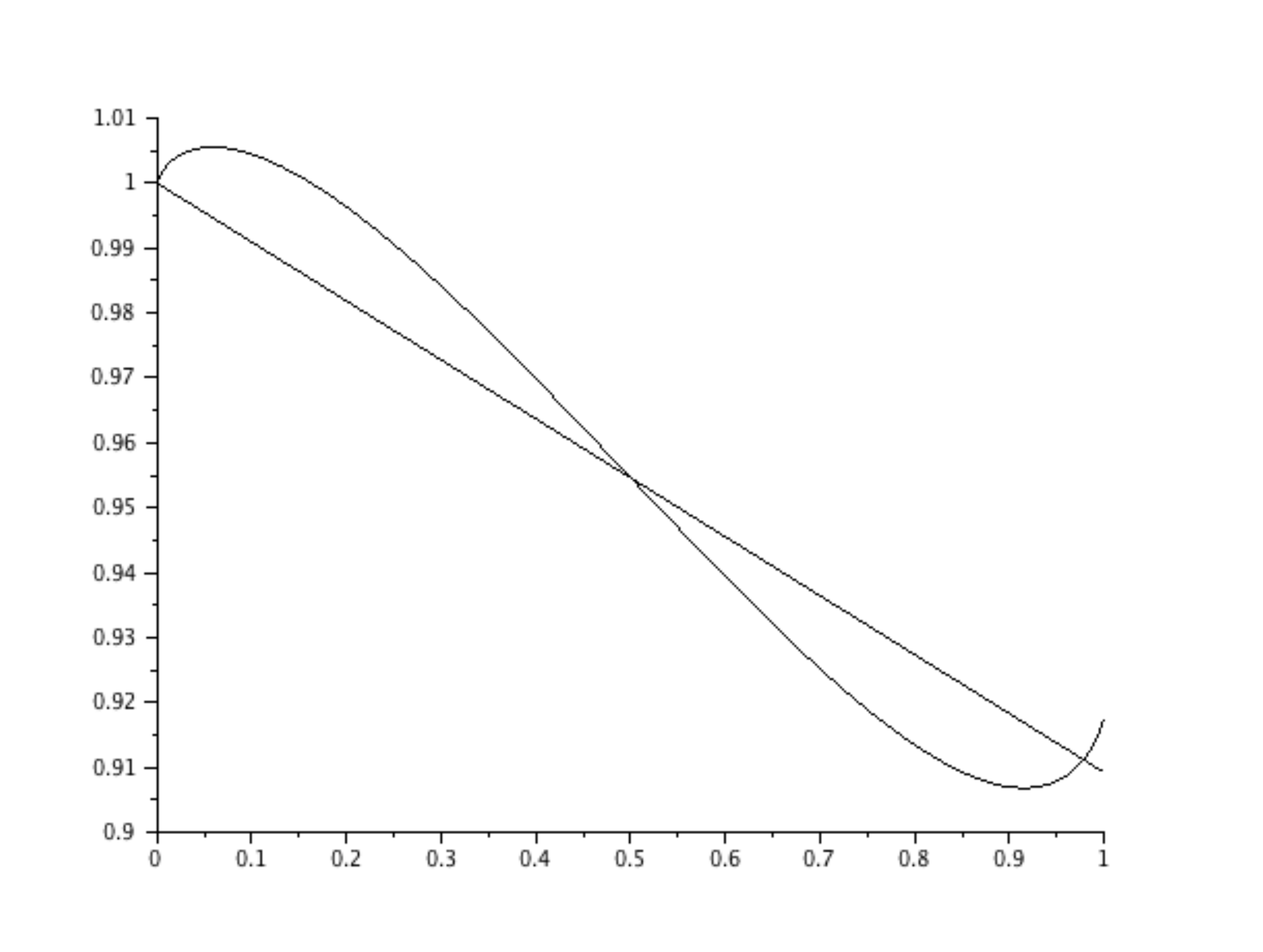}
\caption{Plot of $p(x)\mapsto H(Y) - 0.6 H(X)$ for the asymmetric erasure channel  given in Fig. \ref{figC}. The X-axis is $P(X=0)$. The straight line is drawn to connect the value
of the curve at $P(X=0) = 0$ to that at $P(X=0) = \frac 12$, to visually demonstrate
that this line is not tangent to the curve at $P(X=0) = \frac 12$.}
\end{center}
\end{figure}
The second derivative of the function  $p(x)\mapsto H(Y) - 0.6 H(X)$ at $P(X=0)=\frac 12$ is zero. This validates the fact that $\rho_m^2(X;Y)=0.6$. It is clear that the lower convex envelope of the curve does not pass through $P(X=0)=\frac 12$. The straight line in the figure connects the values of the curve at 0 and $\frac 12$ and clearly demonstrates that the line is not a tangent to the curve.
Thus it is clear in the figure that $\rho_m^2(X;Y)$ is the local-convexity condition and not the condition for being on the convex envelope.
\item Thm. 8 of \cite{AhlswedeGacs} asserts that for fixed $p(y|x),$ $$\max_{p(x)}\rho_m^2(X;Y)=\max_{p(x)}s^*(X;Y).$$
Using the interpretation from Theorem \ref{Thm2} this is immediate since
having a positive semidefinite Hessian at
all points in the domain implies the graph is
convex. Thus, both quantities above equal the minimum value of $\lambda$ such that  
the function $p(x)\mapsto H(Y)-\lambda H(X)$ is convex.
\item The above characterization of $s^*(X;Y)$ is also partly motivated from K\"{o}rner and Marton's characterization in \cite{KorMar76} of less noisy broadcast channels, where they show that for a broadcast channel $X \to (Y,Z)$ the following holds:
$$ I(U;Y) \geq I(U;Z) ~ \forall \ U \to X \to (Y,Z) \iff D(r(z)\|p(z)) \leq  D(r(y)\|p(y))~ \forall r(x),p(x),  $$
where $r(y), r(z)$ are the corresponding channel-induced distributions at $Y$ and $Z$ when $X \sim r(x)$ and similarly $p(y), p(z)$ are the corresponding channel-induced distributions at $Y$ and $Z$ when $X \sim p(x)$.
\end{itemize}

\subsection{Alternate proof for the tensorization of  $s^*(X;Y)$}
The above characterization of  $s^*(X;Y)$ results in an alternate proof of its tensorization.  This proof is directly motivated by the factorization inequalities in broadcast channels, 
some of which can be found in \cite{GengNair}.
Take a distribution of the form $p(x_1, x_2, y_1, y_2)=p_1(x_1)p_1(y_1|x_1)p_2(x_2)p_2(y_2|x_2)$. The easy direction is that $s^*(X_1X_2;Y_1Y_2)\geq \max (s^*(X_1; Y_1), s^*(X_2; Y_2))$. This easily follows from the definition of $s^*(X;Y)$. Thus the non-trivial part is to show that $s^*(X_1X_2;Y_1Y_2)\leq \max (s^*(X_1; Y_1), s^*(X_2; Y_2))$.

Let $\lambda := \max (s^*(X_1; Y_1), s^*(X_2; Y_2))$. With $\mathcal{K}$ denoting the lower convex envelope operator, as earlier, we have
$t_\la(X_1) = \cxe[t_\lambda](X_1)$ at $p_1(x_1)$ and 
$t_\la(X_2) = \cxe[t_\lambda](X_2)$ at $p_2(x_2)$, where 
$t_\la(X_1)$ denotes $H(Y_1)-\lambda H(X_1)$ and 
$t_\la(X_2)$ denotes $H(Y_2)-\lambda H(X_2)$.

We need to show that 
$t_\lambda(X_1,X_2) = \cxe[t_\lambda](X_1,X_2)$ at $p_1(x_1)p_2(x_2)$, 
where $t_\lambda(X_1,X_2)$ denotes 
$H(Y_1, Y_2)-\lambda H(X_1, X_2)$, thought of as a function of 
$p(x_1,x_2)$, with the channel given by 
$p(y_1,y_2|x_1, x_2)=p_1(y_1|x_1)p_2(y_2|x_2)$.

Since for any $W$ satisfying the Markov chain $W-X_1X_2-Y_1Y_2$, 
we have
\begin{align*}H(Y_1, Y_2|W)-\lambda H(X_1, X_2|W)&=H(Y_1|W)-\lambda H(X_1|W)+H(Y_2|W,Y_1)-\lambda H(X_2|W, X_1)
\\&\geq H(Y_1|W)-\lambda H(X_1|W)+H(Y_2|W,Y_1, X_1)-\lambda H(X_2|W, X_1)
\\&= H(Y_1|W)-\lambda H(X_1|W)+H(Y_2|W, X_1)-\lambda H(X_2|W, X_1)~,
\end{align*}
we conclude that
\[
\cxe[t_\la](X_1,X_2) \ge \cxe[t_\la](X_1) + \cxe[t_\lambda](X_2)~.
\]
This inequality in fact holds for all $\lambda$ and for all $p(x_1,x_2)$, not just
for the specific $\lambda$ under consideration and at $p_1(x_1)p_2(x_2)$, which is
where we want to use it.

Now with $(X_1,X_2) \sim p_1(x_1)p_2(x_2)$, we also have
\[
H(Y_1, Y_2)-\lambda H(X_1, X_2) =
H(Y_1)-\lambda H(X_1)+H(Y_2)-\lambda H(X_2)~,
\]
i.e.
we have
$t_\la(X_1,X_2) = t_\la(X_1) + t_\la(X_2)$ at $p_1(x_1)p_2(x_2)$. We can put together
the facts so far to write
\[
t_\la(X_1,X_2) = t_\la(X_1) + t_\la(X_2) = \cxe[t_\la](X_1) + \cxe[t_\la](X_2)
\le \cxe[t_\la](X_1,X_2)~,
\]
holding for the specific $\lambda$ as defined above and for $(X_1,X_2) \sim p_1(x_1)p_2(x_2)$. But by our characterization of $s^*(X_1X_2;Y_1Y_2)$, this
implies that $s^*(X_1X_2;Y_1Y_2)\leq \max\{s^*(X_1; Y_1), s^*(X_2; Y_2)\}$,
completing the proof of the non-trivial direction.

\section{Conclusion}

In this paper we presented a new geometric characterization of the maximal correlation,
$\rho_m(X;Y)$, of a pair of discrete random variables $(X,Y)$ taking values
in finite sets. We also presented a new geometric characterization of
the chordal slope of the nontrivial boundary of the 
hypercontractivity ribbon of $(X,Y)$ at infinity, $s^*(X;Y)$. We showed the application of
these new characterizations in recovering
some of the known results about these quantities in a simple way. We also made a correction
to a data processing inequality claimed by Erkip and Cover \cite{ErkipCover}, the 
error in which has had some knock-on effects in the literature. It would be
interesting to find other connections between the curve
$t_\la(X)$ that we have associated to the channel $p(y|x)$ and the entire 
hypercontractivity ribbon of $(X,Y)$, as we vary $p(x)$.

\begin{center}
ACKNOWLEDGEMENTS
\end{center}

S. Kamath and V. Anantharam gratefully acknowledge research support from the 
ARO MURI grant W911NF-08-1-0233, ``Tools for the Analysis and Design of Complex
Multi-Scale Networks", from the NSF grant CNS-0910702, 
and from the NSF Science \& Technology Center grant CCF-0939370, ``Science of 
Information". The work of Chandra Nair was partially supported by
the following:  an area of excellence  grant (Project No.\ AoE/E-02/08) and two
 GRF grants  (Project Nos. 415810 and 415612)  from the University Grants Committee of the Hong Kong Special Administrative Region, China.

\end{document}